\documentclass[10pt]{bmc_article}    
\usepackage{graphicx}
\usepackage{cite} 
\usepackage{url}  
\usepackage{ifthen}  
\usepackage{multicol}   
\usepackage[latin1]{inputenc} 
\newboolean{publ}

\newenvironment{bmcformat}{\fussy\setboolean{publ}{true}}{\fussy}

\begin{document}
\begin{bmcformat}

\title{Computational identification of transcription factor binding
  sites by functional analysis of sets of genes sharing
  overrepresented upstream motifs}

\author{
        Davide Cor\`a$^1$%
        \email{Davide Cor\`a - cora@to.infn.it}%
      ,
        Ferdinando Di Cunto$^2$%
        \email{Ferdinando Di Cunto - ferdinando.dicunto@unito.it}
      ,
        Paolo Provero$^3$
        \email{Paolo Provero - paolo.provero@fobiotech.org}
      ,
        Lorenzo Silengo$^2$
       \email{Lorenzo Silengo - lorenzo.silengo@unito.it} 	
      and 
        Michele Caselle\correspondingauthor$^1$%
        \email{Michele Caselle\correspondingauthor - caselle@to.infn.it}%
      }

\address{%
    \iid(1) Dipartimento di Fisica Teorica, Universit\`a di Torino, and INFN, sezione di Torino, Italy\\
    \iid(2) Dipartimento di Genetica, Biologia e Biochimica, Universit\`a di Torino, Torino, Italy\\
    \iid(3) Fondazione per le Biotecnologie, Torino, Italy
}%

\maketitle

\begin{abstract}
        \paragraph*{Background:} Transcriptional regulation is a key mechanism in the
	functioning of the cell, and is mostly effected through transcription factors
	binding to specific recognition motifs located upstream of the coding region
	of the regulated gene. The computational identification of such motifs
	is made easier by the fact that they often appear several times in the
	upstream region of the regulated genes, so that the number of
	occurrences of relevant motifs is often significantly larger than
	expected by pure chance.\\

        \paragraph*{Results:} To exploit this fact, we construct sets of
	genes characterized by the statistical overrepresentation of a certain
	motif in their upstream regions. Then we study the functional
	characterization of these sets by analyzing their annotation to Gene
	Ontology terms. For the sets showing a statistically significant
	specific functional characterization, we conjecture that the upstream 
	motif characterizing the set is a binding site for a transcription
	factor involved in the regulation of the genes in the set.\\

        \paragraph*{Conclusions:} The method we propose is able to identify many
	known binding sites in {\it S. cerevisiae} and new candidate targets of
	regulation by known transcritpion factors. Its application to less well
	studied organisms is likely to be valuable in the exploration of their
	regulatory interaction network.

\end{abstract}



\section*{Background}
The regulation of gene expression in the eukaryotic cell happens at 
several different levels, the transcriptional one being among the most
important. The general mechanism is fairly well understood, and
involves the interaction between a trans-acting element, usually a
protein, and a cis-acting element, a recognition site located upstream
of the coding region of the regulated gene and consisting in a rather
short DNA sequence to which the transcription factor is able to bind.  
When bound to the cis-acting elements, the trans-acting ones interfere 
with the transcription machinery, and can either enhance or suppress  
the synthesis of mRNA.

While many instances of this mechanism have been known in great detail
for some time, it is only recently, thanks to the availability of
several fully sequenced genomes and other experimental data on the    
scale of the entire genome, that a study of transcriptional regulation
on a global scale has become possible. Given the sheer size of the
data, the computational aspects of this analysis are highly
non-trivial, and many algorithms have been proposed to select the most
relevant information and exploit it towards a better understanding of
the phenomenon.

One of the most interesting problems is to identify, by purely
computational means, candidate cis-acting elements, so as to choose
promising targets for the experimental investigation and thus greatly
enhance its effectiveness. At first sight this task might seem
prohibitive, since the relevant upstream motifs are rather short 
sequences (in the range of 5 to 20 base pairs) to be found
within hundreds or thousands of base pairs upstream of the coding
region. However, often the relevant motifs must be repeated many  
times, possibly with small variations, in the upstream region for the
regulatory action to be effective. This fact can be exploited to
separate the signal from the noise by searching the upstream region  
for overrepresented motifs, that is motifs appearing many more times
than expected by chance on the basis of suitably chosen background
frequencies.

This strategy was first suggested in \cite{vanhelden:98} where the   
following method was devised to identify regulatory sites: consider a
set of genes experimentally known or presumed to be coregulated (for
example because they are involved in the same biological process or
because they show similar expression profiles in microarray
experiments). Then determine which short motifs are overrepresented in
their upstream region, compared to suitably defined background motif 
frequencies that take into account the basic features of non-coding
DNA of the organism under study. These motifs are likely to be
involved in the coregulation of the genes in the set.

Recently, a different method was proposed by some of us \cite{caselle:02}
which, while also based on the statistical overrepresentation of
regulatory motifs, reverses the procedure compared to \cite{vanhelden:98}
and to most other computational methods: first, the genes are grouped
based on the motifs that are overrepresented in their upstream region;
then the sets of genes thus obtained are analysed from the point of
view of the coregulation of the genes they contain, by studying their
expression patterns in microarray experiments.  The overrepresented   
motifs labelling the sets which do show evidence of coregulation are 
considered as candidate binding sites for the regulation of the genes
in the set. The method was tested on yeast ({\it S. cerevisiae}) and gave
satisfactory results, correctly identifying many previously known
regulatory elements and proposing a few new candidates.

In this work we examine the sets of genes constructed by the same
recipe from a different point of view, namely their functional
characterization based on their annotation to Gene Ontology
\cite{geneontology:00} terms, the rationale being that if a significant
fraction of the genes included in the set corresponding to a given 
motif share the same functional characterization, then it is likely  
that the motif is involved in their coregulation. This procedure for  
validating the sets is likely to be able to recognize different motifs
compared to the validation through microarray data, which is limited 
by at least two factors: first, the specific biological processes and
environmental conditions probed by each microarray experiment, and
second, the fact that only rather large sets of coregulated genes are
likely to produce a statistically significant signal. Therefore it can
be hoped that this analysis of the sets will be able to complement and
improve the one based on expression data.

\section*{Results}
For all possible motifs 5 to 8 nucleotides long we generated the set
of genes in  
whose upstream region the motif is overrepresented compared to its
prevalence in 
the upstream region of all genes. A motif occurring $n$ times in the
upstream region of a gene is considered overrepresented if the probability of
having $n$ or more occurrences by pure chance, computed as described in the
Methods section, is less than 0.01.

\newpage

Given the leniency of the cutoff used to construct the sets, essentially all
possible motifs
turned out to be overrepresented in some genes. It is important to keep in
mind that no biological significance is attached to the sets of genes thus
constructed before they are tested for common functional characterization as
described below. The average number of genes in each set varied
between 11.9 for 
7-letter motifs and 41.8 for 5-letter motifs. 

The actual number of repetitions of each motif in the upstream
sequence needed for a gene to be included in the set greatly varies
depending on the length of the motif and of the upstream sequence and
the background frequency of the motif. For example, 9 repetitions of
the motif AAATT are needed for a 500 bp upstream sequence, but just
one instance of the motif AACCGCGT is enough to make it
overrepresented.  The minimum number of repetitions
for the motifs that passed the functional characterization test is reported in
the supplementary material, together with the number of occurrences of
a sample motif (ACGCG) in the upstream regions of all the genes
included in its set.

We then analyzed the functional characterization of all the sets by studying
the prevalence of genes annotated to each Gene Ontology
term. The statistical significance of such prevalence was evaluated using the
hypergeometric distribution and the false discovery rate was estimated by
comparing the results with the ones obtained for randomly generated gene sets.
With the false discovery rate set at 0.01, a  total
of 107 association were established between 79 different motifs and 41
Gene Ontology terms. These results are shown in Tables 1-6.

\subsection*{Determination of consensus sequences}
The motifs identified by the algorithm clearly
form several groups: within each group the motifs are very similar to
each other and are associated to the same or to very similar Gene
Ontology terms.
For each group of motifs we report a consensus
sequence obtained by aligning (by hand) the motifs corresponding to related
Gene Ontology terms. These consensus sequences are written in terms of
the IUPAC
code with the rules applied in the TRANSFAC database\cite{matys:03}.
Note that the
intrinsic variability
of regulatory binding sites, that is apparently neglected by our method
since we construct the sets using fixed motifs, reappears naturally in
the results when the significant motifs are grouped together.
\par
To verify whether the consensus sequences thus obtained are of
biological relevance, we considered for each of them how many genes
actually contain the consensus sequence in their upstream region,
and what fraction of these are annotated to the GO terms from which
the consensus was obtained.
The results confirm the
correctness of the consensus sequence thus obtained:
for example 17
genes contain the sequence  GATGAGATGAGCT in the upstream region, and
8 of them are annotated ``nucleolus'', while the sequence MCSMTACAY appears
in the upstream region of 78 genes, 20 of which being annotated
``cytosolic ribosome''. The complete results are shown in Tab. 7.
The only case in which the alignment does
not work is the sequence AAMYGCGAWCG, which appears in 6 genes, none
of which is annotated ``nucleosome''. In all cases except the latter
one the genes annotated to the GO term are overrepresented among the
ones containing the consensus sequence in their upstream region, the
most significant P-value being $1.13\cdot 10^{-15}$ for genes
annotated ``ctytosolic ribosome'' and containing the consensus
sequence MCSCMTACAY.
\par
Tables 1-6 show also the transcription factors significantly
associated to each of the motifs found: such association is
determined, as described in the Materials and Methods section, as a
statistically significant intersection between the set of genes
characterized by overrepresentation of the relevant motif and the set
of genes identified in Ref.\cite{lee:02} as able to interact with the
transcription factor. The large fraction of motifs identified by our
method which do have a significant intersection with the sets of genes
identified in Ref.\cite{lee:02} is proof of the effectiveness of our
algorithm, given the completely independent approaches used to define
our sets of genes and the ones of Ref.\cite{lee:02}. Moreover this
analysis allows us to associate known transcription factors to our
candidate motifs, and helps in constructing longer upstream binding
sequences from the motifs we have identified.

\subsection*{The candidate binding sites: detailed discussion}
The two largest groups of motifs (Tab. 1) are associated to
nucleolar proteins implicated in ribosome biogenesis, and can be
easily recognized as the known RRPE \cite{hughes:2000} and PAC
\cite{dequard-chablat:1991} motifs, whose
combinatorial effect in regulating the genes involved in ribosome
biogenesis has been recently described in \cite{sudarsanam:02}.
Ribosomal proteins are associated to a different group of motifs
(Tab. 2). The sets associated to these motifs have significant
intersection with the targets of regulation by the transcription
factors FHL1 and RAP1, and the motifs align well with the binding
sites for these transcription factors as reported in the TRANSFAC database
and in Ref. \cite{lee:02}.
\par
A third large group of motifs displayed a strong
overrepresentation of terms related to DNA replication. For this
group, the consensus sequence perfectly overlaps with one of the
putative Mbp1 binding sites proposed in Ref. \cite{lee:02}.
Interestingly,
together with the regulatory subunit Swi6, this protein forms the MluI
Binding Factor (MBF), a crucial regulator of the G1/S transition
\cite{primig:92}.
A few slightly variant motifs and a completely different one (CGGGGAGA)
are associated  with the annotation "nucleosome".
\par
Besides these large groups, in other four cases the consensus sequence
obtained by merging motifs with similar annotation perfectly overlapped
with the transcription factor binding sites predicted on the basis of
the chromatin IP experiments, as well as with the experimentally
determined sequences as reported in the TRANSFAC database.
In particular, perfect consensus sequences for
GCN4, Ste12, Hsf1 and MET4 were obtained from motifs annotated amino
acid biosynthesis, sexual reproduction, heat shock protein and sulfur
amino acid biosynthesis, respectively (Tab. 4).
\par
Interestingly, in some
instances a perfect match of the surviving motifs with a known binding
site could be obtained by analysis of the literature, even though it
was not reported in Chromatin IP experiments, as shown by the results
reported in Tabs. 5 and 6.  In particular, the sets
characterized by the annotation 'siderophore transport' perfectly
match the binding site for the Aft1p/2p transcription factors,
recently identified as critical regulators of iron uptake
\cite{rutherford:03}. The sets
characterized by the annotation 'tricarboxylic acid cycle' match the
known binding site of the UME6, a pleiotropic regulator implicated in
glucose repression \cite{kratzer:97}.

The sets characterized by the annotation alpha-trealose phosphate synthase
complex  match the Stress Response Element (STRE)
\cite{martinez-pastor:96}, a result consistent with the
described accumulation of trehalose under stress conditions
\cite{winderickx:96}.
\par
The motif
TCGTTTA, associated with the keyword 'steroid metabolism' corresponds
fairly well to the TCGTATA Sterol Response Element (SRE), recently
identified by promoter comparison of the ERG2 and ERG3 genes \cite{vik:01}.
Therefore the genes that are in the corresponding set and are
annotated to ``steroid metabolism'' (listed in Tab. 8) can be
considered as candidate
targets of regulation by UPC2p and ECM22p, that is the transcription
factors identified in Ref.\cite{vik:01} as binding to the SRE
site. Interestingly, ECM22 itself belongs to the set corresponding to
the motif TCGTTTA, hinting to a feedback mechanism in the
regulation of steroid metabolism.
\par
A particularly interesting result is given by the word AAACAAA,
associated with the annotation 'spindle', and in particular with
spindle pole bodies structure and/or function. In this case, no
transcription factors binding to the respective promoters were identified in
Ref. \cite{lee:02}.
However the motif perfectly matches the Hcm1p binding site,
experimentally determined in the promoter of the spindle pole
component SPC110 (WAAYAAACAAW, Ref.\cite{zhu:98}).
Accordingly, Hcm1p has been implicated, by a different ChIP
analysis \cite{horak:02}, as
one of the potential regulators of spindle pole body genes, even
though 8 of the 9 genes (the exception is CIN8) identified by our
approach and listed in Tab. 9 are not included in the list of the 262
potential Hcm1p targets. More precisely Tab. 9 lists the genes that
are included in the set corresponding to the motif AAACAAA and are
annotated to the term "spindle". Our results strongly suggest that
these genes are potential targets of the Hcm1 transcription factor.

Finally the consensus sequence derived from motifs with a significant
overrepresentation of 'proteasome' perfectly corresponds to the
proteasomal element identified in Ref. \cite{jensen:00} through a different
computational approach, to which no transcription factors have been
experimentally associated.

\subsection*{Functional characterization of the sets corresponding to
  candidate binding sites}

The statistical significance of the associations we made between
motifs and Gene Ontology terms is based on the {\it overrepresentation} of
genes annotated to the term in the set of genes defined by the motif:
clearly not all genes in the various sets are annotated to the term
itself. It is therefore interesting to examine the annotation of the
genes in the set that are not annotated to the relevant term. To
perform this analysis in a systematic way, we defined, for each Gene
Ontology term, three categories of genes:
\begin{itemize}
\item
  Genes {\it annotated} to the term, {\it i.e.} directly annotated to
  the term itself or to any of its descendants in the Gene Ontology
  graph
\item
  Genes whose annotation is {\it compatible} with the term, defined as
  genes that are either directly annotated to an ancestor of the term,
  or to ``cellular component unknown'', etc. or finally genes which do
  not have any annotation in the relevant branch of the Gene Ontology
\item
  All other genes are considered {\it incompatible} with the Gene
  Ontology term
\end{itemize}
For example if the term at hand is ``nucleolus'', a gene annotated
``nucleus'' is considered compatible, while a gene annotated
``cytoplasm'' is incompatible.

A few examples of the results of this analysis: 
the set defined by the CCATACA motif includes a total of
19 genes: 10 of them are annotated ``ribosome biogenesis'', 2 are
compatible with such annotation and the remaining 7 are
incompatible. The AAACAAA set includes 55 genes: 9 of them are 
annotated ``spindle'', 31 are compatible with such annotation and 15
are incompatible. The complete results are available in the
supplementary material.

\subsection*{Localization of the candidate binding sites with respect
  to the Transcription Start Site}

It is interesting to investigate possible correlations between the location
 of the (candidate) binding sites that we find and the Transcription Starting
 Site (TSS in the following) of the genes. Ideally one would expect these
 binding sites to be preferentially located upstream with respect to the TSS
 and clustered in the first 100-200 bases upstream the TSS. This test however is
 rather non trivial due to the lack of systematic knowledge of TSS's. Among the
 possible sources of information on the TSS's we decided to use  UTRDB
 \cite{pesole:99}
 which contains  5' UTR sequences for 330 yeast genes.
 We looked at the intersection between these 330 gene and our sets for the
 candidate binding sites listed in Tab. 1-6. We selected for a deeper analysis
 the intesection corresponding to the consensus sequence GATGAGATGAGCT. The
 intersection contains 6 genes which are shown in Fig. 1 and a total of 18
 instances of the candidate binding sites. As it can be seen from the figure the
 majority (66\%) of the binding sites are located upstream of the TSS and half
of
 them are within 200 bases from the TSS. However
 this a slightly misleading indicator since some of the UTR's are very short in
 length. If we look instead at the density of binding sites, i.e. at the ratio
 between the number of instances and the length of the observed sequence, we
 find a slightly higher density in the UTR regions (about 1 instance every 100
 bases) with respect to the truly upstream regions (1 instance every 150 bases).
 We plan to address this subject in a more complete way in a
 future work.

\subsection*{Control analyses}

Recently a major revision of the list of yeast ORFs, and in particular
the elimination of many putative ORFs, was suggested in
Ref.\cite{kellis:03} through comparison with closely related
organisms. To verify whether such changes in the annotation of the
genome had an important effect on our results, we repeated our
analysis without the $\sim 500$ ORFs that Ref\cite{kellis:03} suggests
to eliminate; the results however did not change in any significant
way (data not shown), clearly because most of the ORFs to be eliminated are
not annotated to Gene Ontology terms.

\newpage

Finally, as a general check of the soundness of this procedure, we
repeated the whole analysis on sequences which are not assumed to be
as relevant to transcriptional regulation as the upstream sequences we
have studied in depth. More precisely, we selected all yeast intergenic
regions that are not upstream of any gene and we associated each
region to the genes with respect to which the sequence is located
downstream. Then we repeated the whole analysis in precisely the same
way as for the upstream regions: no significant results were found with
a false discovery rate of 0.01.

\section*{Discussion}
The technique we have proposed in this paper is able to identify many
binding sites related to functional groups of different size. The
binding sites we identified were all previously  known from different methods
(in some cases from computational analyses only).
This is explained partly by the fact that yeast is among the
best-known organisms for what concerns transcriptional regulation, and
partly by our very restrictive choice of false discovery rate (1\%).
Indeed the main goal of our analysis was to show that the method,
and in particular the statistical analysis used to establish the
significance of the results, is reliable, and can therefore be used on
less well known organisms where it is likely to produce new candidate
binding sites.

Besides identifying new binding sites, our method could be extremely
valuable in identifying new potential targets of known transcription
factors. Indeed, in the case of Hcm1p we were able to identify 8 new
potential target genes that were not identified in chromatin
immunoprecipitation experiments.

An important aspect of our technique is that the gene sets are based
exclusively on the upstream sequence. Therefore the same sets can be
validated from the biological point of view in many different ways, of
which the use of microarray results, as in Ref.\cite{caselle:02}, of Gene
Ontology annotations and of the results of chromatin
immunoprecipitation experiments (as in the present paper) are just the
most natural examples. These different validation methods can reveal
different binding sites and should be combined whenever possible: for
example of all the binding sites revealed
in the present analysis, only the ones pertaining to the regulation of
ribosome biogenesis and the stress response element CCCCT had been previously
identified by the microarray validation of
Ref. \cite{caselle:02}. Conversely, the AGCCGCGC binding site of the UME6
transcription factor, correctly identified in Ref.\cite{caselle:02},
was not found in the present one.

Another possible development of this line of research is the
construction of sets of genes labeled by spaced
dyads instead of simple sequences \cite{vanhelden:00}. These sets
could then be validated either based on the overrepresentation of Gene
Ontology annotations, as in the present work, or based on their
coexpression as determined in microarray experiments as in
Ref.\cite{caselle:02}.

\section*{Conclusions}
We have presented a new algorithm for the identification by
computational means of transcription factor binding sites. The
algorithm proceeds by first defining sets of genes having in common
the overrepresentation of a certain short DNA sequence in their
upstream region, and then selecting among all these sets the ones
showing a statistically significant overrepresentation of genes
annotated to specific Gene Ontology terms.

When applied to yeast the algorithm is able to independently find many
known transcription factor binding sites. It is
therefore likely that our algorithm could be profitably used in the
study of transcripional regulation in higher eukaryotes and in humans
in particular. Here the method needs to be modified, to take into
account the increased complexity of the problem: the strategy we are
currently following is to limit the analysis to those upstream regions
that are highly conserved in mouse and human. These conserved regions
are likely to contain many binding sites, and are short enough to
provide a suitable signal to noise ratio. The results will be
presented in a forthcoming publication.

While most regulatory binding sites are found upstream of the coding sequence,
regulatory regions can often be found, for example, in the 3' region
or within the first intron: the method we have presented can in
principle be applied, with suitable modifications, to the
investigation of such regions. Moreover, similar methods could be
devised for the computational identification of other functionally
relevant DNA sites.

\section*{Methods}
\subsection*{Upstream sequences}
For all yeast open reading frames (ORFs) we
considered up to 500 bp upstream of the translation start, cutting the
sequence shorter when necessary to prevent overlapping with the coding
region of neighboring ORFs. The sequences were obtained from the
Regulatory Sequence Analysis Tool \cite{vanhelden:03}.

\subsection*{Purging repeated upstream sequences}
Genes with highly similar
upstream regions are a likely source of false positives, especially
when similarly annotated. Consider for example the ASP3 gene, which
appears in four copies in the genome, with nearly identical upstream
regions, and identical Gene Ontology annotations. Clearly these four
upstream regions must be counted as one to avoid introducing a bias
that is likely to generate false positives. To avoid these problems,
we first used BLAST \cite{altschul:90} to list all the pairs of nearly
identical upstream sequences (Blast P-value less than
$10^{-90}$). Then we used these results to form clusters of
nearly identical upstream regions (that is the connected components of
the graph constructed  by joining with an edge all pairs of nearly
identical upstream sequences). Finally, we retained for further
analysis only one gene per group, chosen at random.  After this
procedure we were left with 6,037 upstream sequences which we retained
for further analysis. 

\subsection*{Motifs analyzed} 
We considered all possible DNA sequences 5 to 8 base
pairs long.  Occurrences of each motif were counted on both strands:
therefore, for example, the motifs CCCCT and AGGGG are considered as
the same motif, and the number of occurrences of this motif is the sum
of the number of occurrences of CCCCT and AGGGG on one strand. For
palindrome motifs (such as AGGCCT) the number of occurrences is simply
the number of occurrences on one strand. 

\subsection*{Background frequencies}  
The determination of which motifs are
overrepresented in each upstream region needs the definition of a
background frequency to which the number of actual occurrences is
compared. We chose to use the frequency of each motif in the set of 
all the upstream sequences considered in the study, taken as a single
sample. Denote by $U(g)$ the length of the upstream region considered
for a gene $g$ (500 bp or less in this study, see above), and by $n(m,g)$
the number of occurrences of the motif $m$ in such region. If $b(m)$ is
the length of $m$, the background frequency $f(m)$ of $m$ is defined as 

\begin{equation}
f(m) = \frac{\sum_g n(m,g)}{\sum_g u(g,m)}
\end{equation}
where both sums are taken over all the yeast ORFs. 
In the denominator
$$u(g,m) \equiv U(g) - b(m) + 1$$
is  the number of words of length $b(m)$ that can be read in the upstream region 
of $g$.

\subsection*{Overrepresented motifs}
The null hypothesis we use to determine
whether a motif is overrepresented in the upstream region of a gene is
the same as in Refs. \cite{vanhelden:98} and \cite{caselle:02}: motifs are
distributed randomly, each with its own background frequency, in the
upstream regions. The probability of having $n(m,g)$ or more occurrences
of $m$ in the upstream region of $g$ depends on the background frequency
of $m$ and the length $U(g)$ of the upstream region of $g$, and  is given by
the right tail of the binomial distribution:
\begin{eqnarray}
P(n(m,g),f(m),U(g))=\sum_{k=n(m,g)}^{u(g,m)}
\left({u(g,m) \atop k}\right) \left[f(m)\right]^{k}
\left[1-f(m)\right]^{u(g,m)-k}\ \
\label{overrep}
\end{eqnarray}
Notice that the binomial distribution is based on the assumption that
the motifs read successively in the upstream sequence
are independent of each other; while this is
obviously not the case, the motifs we study (5 base pairs or longer)
are sufficiently rare that this aspect can be safely neglected.
\par
The motif $m$ is considered overrepresented in the upstream region of $g$ if
the probability 
$$P(n(m,g),f(m),U(g))$$ 
is lower than a certain cutoff. 
In this study we fixed the cutoff at 0.01. It is important to
keep in mind that no
biological significance is attributed to this overrepresentation by
itself: only the motifs that pass the functional characterization test
described below will be retained in the final results. Therefore the
choice of this cutoff can be arbitrarily lenient. While it
  would seem natural to make the cutoff more strict as the word length
  is increased, to take into account the larger number of words
  analysed, our experience, also gained with the microarray analysis
  of reference \cite{caselle:02} shows that many significant results
  would be lost in this way.
The result of this
analysis is, for each motif $m$, a set $S(m)$ of genes in whose upstream
region $m$ is overrepresented.

\subsection*{Functional characterization of the sets $S(m)$}
The final step is the
analysis of the annotation of the genes included in each set $S(m)$, to
select the sets characterized by a strong functional characterization
and hence the candidate binding sites. For each set $S(m)$ we computed
the prevalence of all Gene Ontology (GO) terms among the annotated
genes in the set, and the probability that such prevalence would occur
in a randomly chosen set of genes of the same size. We always consider
a gene annotated to a GO term if it is directly annotated to it or to any
of its descendants in the GO graph. For a given GO term $t$ let $K(t)$ be
the total number of ORFs annotated to it in the genome, and $k(m,t)$ the
number of ORFs annotated to it in the set $S(m)$. If $J$ and $j(m)$ denote
the number of ORFs in the genome and in $S(m)$ respectively, such
probability is given by the right tail of the appropriate
hypergeometric distribution:
\begin{equation}
P(J,K(t),j(m),k(m,t)) = \sum_{h=k(m,t)}^{{\rm min}(j(m), K(t))} F(J,
K(t), j(m), h)
\end{equation}
where
\begin{equation}
F(M,m,N,n) = \frac{\left(m \atop n\right) \left(M - m \atop N -
  n\right)}{\left(M \atop N\right)}
\end{equation}
In this way a P-value can be associated to each pair made of a motif
and a Gene Ontology term. A low P-value indicates that the
overrepresentation of the motif is correlated to the functional
characterization described by the GO term, and hence suggests the
motif as a candidate binding site for the genes with such functional
characterization.

\subsection*{Determining the cutoff on P-values}
Given the huge number of P-values
that we compute (in principle equal to the number of GO terms
multiplied by the number of motifs analysed) it is clear that
very low P-values could appear simply by chance. Therefore a
careful analysis is required to keep under control the number
of false positives. The usual way of dealing with this issue,
that is the Bonferroni correction, is not appropriate, because
due to the hierarchical nature of the Gene Ontology annotation
scheme, the P-values we compute are very far from being
independent from each other (for example, for a given motif,
the P-values associated to terms such as 'cell cycle and DNA
replication' and 'chromosome cycle' are obviously strongly
correlated to each other). We decided therefore to adopt an
empirical approach allowing us to fix the false discovery rate
without any prior assumption regarding the distribution of the
P-values. 

First, we randomly generated a large number $N_R$ of
sets of genes comparable in size to the typical size of the
sets associated to the motifs. For each of these sets we
computed the P-values associated to all the GO terms with the
same formula used for the "true" sets. For each random set we
considered only the best P-value obtained for each of the
three branches of the Gene Ontology, and we ranked the random
sets based on this best P-value. In this way we can determine
a false discovery probability $p_f(C)$ as a function of the
cutoff on P-values $C$, defined as  
\begin{equation}
p_f(C)=\frac{n_f(C)}{N_R} 
\end{equation}
where $n_f(C)$ is the number of random sets whose best P-value is less
than C. 
\par
Equipped with the function $p_f(C)$, we can estimate the ratio of
false discoveries among our true sets, as a function of $C$: if $N$ is the
total number of true sets examined, the expected number of false
discoveries with cutoff $C$ is  $N p_f(C)$. Therefore if $n(C)$ is the number
of true sets with best P-value less than $C$ we can estimate the false
discovery rate (FDR) to be 
$$\frac{N p_f(C)}{n(C)}\ .$$
In this way we can choose $C$
based on the desired false discovery rate. Clearly the lower is the
FDR required, the higher is the precision required in determining the
function $p_f(C)$ for low values of $C$, and hence the number $N_R$ of sets
to be generated randomly. In our case we set the false discovery rate
at 0.01: a reliable estimate of the corresponding cutoffs in P-value
required the generation of 3.5 million randomly chosen sets of 20
genes (which is the typical size for the true sets corresponding to
overrepresented motifs 5 to 8 bp long). Simulations performed with
other choices of the set size did not differ enough to change significantly
our estimate of the false discovery rate.

\subsection*{Comparison with experimentaly determined associations
  between transcription factors and regulated genes}
We performed a systematic
comparison between the sets of genes identified as significantly
associated to one or more GO terms and the sets of genes that were
determined in Ref.\cite{lee:02} to be capable of interacting with each
of 106 known transcription factors. For each of these transcription
factors we defined a set of regulated genes (called "TF set" in the
following) by  considering all the genes with P-value less than 0.001
for interaction with the TF according to the supplementary material of
Ref. \cite{lee:02}. Then for each of the sets identified as
significant by our method we considered its intersection with all the
TF sets, and computed a related P-value: the probability that a
randomly chosen set of genes has an intersection equal to or greater
(in number of elements) than the one actually found (given by the same
hypergeometric distribution used to determine the overrepresentation
of GO annotations). We considered the result statistically significant
when this intersection P-value was less than $10^{-5}$.

\section*{Authors contributions}
   \ifthenelse{\boolean{publ}}{\small}{}
D.C. implemented most of the algorithms and also partecipated in the project 
design. F.D.C. performed the literature mining for the assignement of the identified
putative binding sites to known transcription factors. P.P. contributed to the
algorithm implementation and wrote the manuscript. L.S. supervised the analysis 
concerning the biological significance of the putative binding sites.
M.C. coordinated the whole project. \\
All the authors have read and approved the final manuscript.

\section*{Acknowledgements}
  \ifthenelse{\boolean{publ}}{\small}{}
We thank Manuela Helmer-Citterich, Carl Herrmann and Maurizio
Pellegrino for inspiring discussions. We are also grateful to the
anonymous referees whose comments and suggestions allowed a
significant improvement of the work.


{\ifthenelse{\boolean{publ}}{\footnotesize}{\small}
 \bibliographystyle{bmc_article}  
  \bibliography{bmc_article} }     




\section*{Figures}
  \subsection*{Figure 1 legend}
Location of the motifs belonging to the consensus GATGAGATGAGCT with
respect to the translation and transcription start sites for 6 genes
for which the latter is known. The binding sites are denoted by rectangles
above or below the line depending on whether the consensus sequence
is read on the Crick or Watson strand respectively. The vertical bar
is the transcription start site, as given in Ref.[20].

\section*{Additional Files}
  \subsection*{Additional file 1 --- overrepresentation.txt}
For all sets found significant by the algorithm this table shows the
background frequency of the motif, and the number of instances
required for the motif to be overrepresented in a 500 bp long
upstream sequence. For shorter upstream sequences less repetitions
are needed in general.

  \subsection*{Additional file 2 --- occurrences.txt}
The number of occurrences of the motif ACGCG (associated to
``replication fork'' and ``DNA metabolism'') in all the genes
belonging to its set, and the length of the corresponding upstream region.

  \subsection*{Additional file 3 --- annotation.txt}
For all sets found significant by the algorithm the table reports the
corresponding Gene Ontology term and the number of genes annotated
to the term, compatible and incompatible with the term (see text
for the definition of "compatible" and "incompatible").

\newpage\noindent
\section*{Tables}
 
\subsection*{Table 1}
 Significant motifs associated to nucleolar proteins
  implicated in ribosome biogenesis. The columns titled ``C'', ``F''
  and ``P'' correspond to the three branches of the Gene Ontology:
  cellular component, molecular function and biological process respectively.  \par \mbox{}
\begin{table}[h]
\begin{center}
\begin{tabular}{|c@{}c@{}c@{}c@{}c@{}c@{}c@{}c@{}c@{}c@{}c@{}c@{}c@{}c@{}|c|c|c|}
\hline
\multicolumn{14}{|c|}{motif} &C&F&P\\
\hline
G&A&T&G&A&G&A& & & & & & &\  &nucleolus&-&ribosome biogenesis\\
\hline
G&A&T&G&A&G&A&T& & & & & & &nucleolus&-&ribosome biogenesis\\
\hline
 &A&T&G&A&G&A&T& & & & & & &nucleolus&-&ribosome biogenesis\\
\hline
 &A&T&G&A&G&A&T&G& & & & & &    -    &-&ribosome biogenesis\\
\hline
 & &T&G&A&G&A&T&G& & & & & &    -    &-&ribosome biogenesis\\
 & & & & & & & & & & & & $\ $& &&& and assembly\\
\hline
 & &T&G&A&G&A&T&G&A& & & & &    -    &-&ribosome biogenesis\\
& & & & & & & & & & & & $\ $& &&& and assembly\\
\hline
 & & &G&A&G&A&T&G& & & & & &    -    &-&ribosome biogenesis\\
& & & & & & & & & & & & $\ $& &&& and assembly\\
\hline
 & & &G&A&G&A&T&G&A&G& & & &nucleolus&-&ribosome biogenesis \\
& & & & & & & & & & & & $\ $& &&& and assembly\\
\hline
 & & &G&A&G&A&T&G&A& & & & &nucleolus&-&ribosome biogenesis \\
& & & & & & & & & & & & $\ $& &&& and assembly\\
\hline
 & & & &A&G&A&T&G&A&G& & & &nucleolus&-&ribosome biogenesis\\
\hline
 & & & & &G&A&T&G&A&G& & & &nucleolus&-&ribosome biogenesis\\
\hline
 & & & & &G&A&T&G&A& & & & &    -    &-&ribosome biogenesis\\
\hline
 & & & & & &A&T&G&A&G&C&T$\ $& &nucleolus&-&ribosome biogenesis\\
\hline
 & & & & & & &T&G&A&G&C&T$\ $& &nucleolus&-&rRNA processing\\
\hline
{\bf G}&{\bf A}&{\bf T}&{\bf G}&{\bf A}&{\bf G}&{\bf A}&{\bf T}&{\bf
G}&{\bf A}&{\bf G}&{\bf C}&{\bf T}&&
\multicolumn{3}{c|}{}\\
\hline
A&A&A&A&A&T&T& & & & & & $\ $& &nucleolus&-&ribosome biogenesis\\
\hline
A&A&A&A&A&T&T&T& & & & & $\ $& &nucleolus&-&transcription\\
 & & & & & & & & & & & & $\ $& &complex&& from Pol I promoter\\
\hline
 &A&A&A&A&T&T& & & & & & $\ $& &nucleolus&-&ribosome biogenesis\\
\hline
 &A&A&A&A&T&T&T& & & & & $\ $& &nucleolus&-&ribosome biogenesis\\
\hline
 &A&A&A&A&T&T&T&T& & & & $\ $& &nucleolus&-&ribosome biogenesis\\
\hline
 & &A&A&A&T&T& & & & & & $\ $& &nucleolus&-&35S primary \\
 & & & & & & & & & & & & $\ $& &&&transcript processing\\
\hline
 & &A&A&A&T&T&T&T&C& & & $\ $& &small nucleolar&-&35S primary\\
 & & & & & & & & & & & & $\ $& &ribonucleoprotein&&transcript processing\\
 & & & & & & & & & & & & $\ $& &complex&&\\
\hline
{\bf A}&{\bf A}&{\bf A}&{\bf A}&{\bf A}&{\bf T}&{\bf T}&{\bf T}&{\bf
T}&{\bf C}&&&&&
\multicolumn{3}{c|}{}\\
\hline

\end{tabular}
\end{center}

\end{table}

\newpage\noindent
\subsection*{Table 2}
Significant motifs associated to ribosomal proteins. 
Here and in the following tables the column "TF" reports the transcription
factors studied in Ref.\cite{lee:02} for which the intersection
between the experimentally found targets of regulation and our sets is
statistically significant. \par \mbox{}
\begin{table}[h]
\begin{center}
\begin{tabular}{|c@{}c@{}c@{}c@{}c@{}c@{}c@{}c@{}c@{}c@{}c@{}|c|c|c|c|}
\hline
\multicolumn{11}{|c|}{motif} &C&F&P&TF\\
\hline
A&C&C&C&A&T&A& & & &$\ $ &cytosolic&structural&-&FHL1\\
 & & & & & & & & & &\  &ribosome &constituent&&RAP1\\
 & & & & & & & & & &\  &(sensu Eukarya)&of ribosome&&\\
\hline
A&C&C&C&A&T&G&C& & &\  &cytosolic&-&-&FHL1\\
 & & & & & & & & & &\  &ribosome &&&RAP1\\
 & & & & & & & & & &\  &(sensu Eukarya)&&&\\
\hline
A&C&C&C&G&T&A&C& & &\  &cytosolic&-&-&FHL1\\
 & & & & & & & & & &\  &ribosome &&&RAP1\\
 & & & & & & & & & &\  &(sensu Eukarya)&&&\\
\hline
C&C&G&C&C&T&A&C& & &\  &cytosolic&-&-&FHL1\\
 & & & & & & & & & &\  &ribosome &&&RAP1\\
 & & & & & & & & & &\  &(sensu Eukarya)&&&\\
\hline
C&C&G&C&C&T& & & & &\  &large&-&-&FHL1\\
 & & & & & & & & & &\  &ribosomal &&&\\
 & & & & & & & & & &\  &subunit&&&\\
\hline
 &C&C&C&G&T&A&C&A& &\  &cytosolic&-&-&FHL1\\
 & & & & & & & & & &\  &ribosome &&&RAP1\\
 & & & & & & & & & &\  &(sensu Eukarya)&&&\\
\hline
 & &C&C&A&T&A&C&A&T&\  &cytosolic&structural&-&FHL1\\
 & & & & & & & & & &\  &ribosome &constituent&&\\
 & & & & & & & & & &\  &(sensu Eukarya)&of ribosome&&\\
\hline
 & &C&C&A&T&A&C&A& &\  &cytosolic&structural&protein&FHL1\\
 & & & & & & & & & &\  &ribosome &constituent&biosynthesis&RAP1\\
 & & & & & & & & & &\  &(sensu Eukarya)&of ribosome&&\\
\hline
 & &G&C&C&T&A&G& & &\  &cytosolic&-&-&FHL1\\
 & & & & & & & & & &\  &ribosome &&&\\
 & & & & & & & & & &\  &(sensu Eukarya)&&&\\
\hline
 & &G&C&C&T&A&G&A&C&\  &cytosolic&-&-&FHL1\\
 & & & & & & & & & &\  &ribosome &&&RAP1\\
 & & & & & & & & & &\  &(sensu Eukarya)&&&\\
\hline
 & &G&C&C&C&A & & & &\  &cytosol&-&-&FHL1\\
 & & & & & & & & & &\  &&&&RAP1\\
\hline
 & & &C&A&T&A&C&A&T&\  &cytosolic&-&-&FHL1\\
 & & & & & & & & & &\  &ribosome &&&\\
 & & & & & & & & & &\  &(sensu Eukarya)&&&\\
\hline
{\bf M}&{\bf C}&{\bf S}&{\bf C}&{\bf M}&{\bf T}&{\bf A}&{\bf C}&{\bf
A}&{\bf Y}&\multicolumn{5}{c|}{}\\
\hline
\end{tabular}
\end{center}
\end{table}

\newpage\noindent
\subsection*{Table 3}
Significant motifs associated to DNA replication and nucleosome. \par \mbox{}
\begin{table}[h]
\begin{center}
\begin{tabular}{|c@{}c@{}c@{}c@{}c@{}c@{}c@{}c@{}c@{}c@{}c@{}c@{}|c|c|c|c|}
\hline
\multicolumn{12}{|c|}{motif} &C&F&P&TF\\
\hline
A&A&A&C&G&C&G& & & & &$\ $&-&-&DNA replication&MBP1\\
 & & & & & & & & & & &\  &&&&SWI6\\
\hline
A&G&A&C&G&C&G&T& & & &$\ $&-&-&DNA dependent&\\
 & & & & & & & & & & &\  &&&DNA replication&\\
\hline
 &G&A&C&G&C&G&T&A& & &$\ $&-&-&DNA replication and&MBP1\\
 & & & & & & & & & & &\  &&&chromosome cycle&SWI6\\
\hline
 &A&A&C&G&C&G& & & & &$\ $&-&-&DNA replication&MBP1\\
 & & & & & & & & & & &\  &&&&SWI6\\
\hline
 &G&A&C&G&C&G& & & & &$\ $&-&-&mitotic sister&MBP1\\
 & & & & & & & & & & &\  &&&chromatid&SWI4\\
 & & & & & & & & & & &\  &&&cohesion&SWI6\\
\hline
 & &A&C&G&C&G& & & & &$\ $&replication fork&-&DNA metabolism&MBP1\\
 & & & & & & & & & & &\  &&&&SWI4\\
 & & & & & & & & & & &\  &&&&SWI6\\
\hline
 & &A&C&G&C&G&T& & & &$\ $&-&-&DNA replication&MBP1\\
 & & & & & & & & & & &\  &&&&SWI6\\
\hline
 & &A&C&G&C&G&T&C&G&&$\ $&-&-&mitotic sister&MBP1\\
 & & & & & & & & & & &\  &&&chromatid&SWI6\\
 & & & & & & & & & & &\  &&&cohesion&\\
\hline
 & & &C&G&C&G&T&A& & &$\ $&-&-&DNA metabolism&MBP1\\
\hline
{\bf A}&{\bf R}&{\bf A}&{\bf C}&{\bf G}&{\bf C}&{\bf G}&{\bf T}&{\bf M}&{\bf
G}&&\multicolumn{5}{c|}{}\\
\hline
A&A&C&C&G&C&G&T& & & &$\ $&nucleosome&-&-&HIR1\\
 & & & & & & & & & & &\  &&&&HIR2\\
\hline
 &A&A&T&G&C&G&A& & & &$\ $&nucleosome&-&-&HIR1\\
 & & & & & & & & & & &\  &&&&HIR2\\
\hline
 & &A&T&G&C&G&A&A& & &$\ $&nucleosome&-&chromatin&HIR1\\
 & & & & & & & & & & &\  &&&assembly/disassembly&HIR2\\
\hline
 & & &C&G&C&G&A& & & &$\ $&nucleosome&-&-&MBP1\\
 & & & & & & & & & & &\  &&&&SWI4\\
 & & & & & & & & & & &\  &&&&SWI6\\
\hline
 & & & &G&C&G&C&T&C&G&$\ $&nucleosome&-&-&\\
\hline
{\bf A}&{\bf A}&{\bf M}&{\bf Y}&{\bf G}&{\bf C}&{\bf G}&{\bf A}&{\bf W}&{\bf
C}&{\bf
G}&\multicolumn{5}{c|}{}\\
\hline
C&G&G&G&G&A&G&A& & & &$\ $&nucleosome&-&-&\\
\hline
\end{tabular}
\end{center}

\end{table}

\newpage\noindent
\subsection*{Table 4}
Other motifs with significant intersection with ChIP data. \par \mbox{}
\begin{table}[h]
\begin{center}
\begin{tabular}{|c@{}c@{}c@{}c@{}c@{}c@{}c@{}c@{}c@{}c@{}c@{}|c|c|c|c|}
\hline
\multicolumn{11}{|c|}{motif} &C&F&P&TF\\
\hline
G&A&T&G&A&G&T&C& & &$\ $&-&-&amino acid&GCN4\\
 & & & & & & & & & &\  &&&metabolism&\\
\hline
 &A&T&G&A&C&T& & & &$\ $&-&-&non-protein&GCN4\\
 & & & & & & & & & &\  &&&amino acid&\\
 & & & & & & & & & &\  &&&metabolism&\\
\hline
 &G&T&G&A&G&T&C&A& &$\ $&-&-&amino acid&GCN4\\
 & & & & & & & & & &\  &&&metabolism&\\
\hline
 &A&T&G&A&G&T&C&A& &$\ $&-&-&amino acid&GCN4\\
 & & & & & & & & & &\  &&&metabolism&\\
\hline
 & &T&G&A&G&T&C&A&C&$\ $&-&-&amino acid&GCN4\\
 & & & & & & & & & &\  &&&biosynthesis&\\
\hline
 & &A&G&A&G&T&C&A&T&$\ $&-&-&amino acid&\\
 & & & & & & & & & &\  &&&metabolism&\\
\hline
 & & &G&A&G&T&C&A& &$\ $&-&-&amino acid&BAS1\\
 & & & & & & & & & &\  &&&biosynthesis&GCN4\\
\hline
{\bf G}&{\bf A}&{\bf T}&{\bf G}&{\bf A}&{\bf G}&{\bf T}&{\bf C}&{\bf A}&{\bf
Y}&\multicolumn{5}{c|}{}\\
\hline
T&G&A&A&A&C& & & & &$\ $&-&-&sexual reproduction&DIG1\\
 & & & & & & & & & &\  &&&&STE12\\
\hline
T&G&A&A&A&C&A& & & &$\ $&-&-&sexual reproduction&DIG1\\
 & & & & & & & & & &\  &&&&STE12\\
\hline
{\bf T}&{\bf G}&{\bf A}&{\bf A}&{\bf A}&{\bf C}&{\bf
A}&&&&\multicolumn{5}{c|}{}\\
\hline
A&C&T&G&T&G& & & & &$\ $&-&-&sulfur amino&MET4\\
 & & & & & & & & & &\  &&&acid transport&\\
\hline
 & &T&G&T&G&G&C& & &$\ $&-&-&sulfur metabolism&MET4\\
 & & & & & & & & & &\  &&&&MET31\\
\hline
{\bf A}&{\bf C}&{\bf T}&{\bf G}&{\bf T}&{\bf G}&{\bf G}&{\bf
C}&&&\multicolumn{5}{c|}{}\\
\hline
T&C&T&A&G&A&A& & & &$\ $&-&heat shock protein&protein folding&HSF1\\
\hline
\end{tabular}
\end{center}

\end{table}

\newpage\noindent
\subsection*{Table 5}
Motifs associated to siderophore transport and
  tricarboxylic acid cycle. \par \mbox{}
\begin{table}[h]
\begin{center}
\begin{tabular}{|c@{}c@{}c@{}c@{}c@{}c@{}c@{}c@{}c@{}c@{}c@{}|c|c|c|}
\hline
\multicolumn{11}{|c|}{motif} &C&F&P\\
\hline
A&G&G&G&T&G&C& & & &$\ $&-&-&siderophore\\
 & & & & & & & & & &\   &&&transport\\
\hline
A&G&G&G&T&G&C&A& & &$\ $&-&-&siderophore\\
 & & & & & & & & & &\   &&&transport\\
\hline
T&G&G&G&T&G&C&A& & &$\ $&-&-&siderophore\\
 & & & & & & & & & &\   &&&transport\\
\hline
 &G&G&G&T&G&C&A& & &$\ $&-&-&siderophore\\
 & & & & & & & & & &\   &&&transport\\
\hline
 &G&G&G&T&G&C& & & &$\ $&-&-&siderophore\\
 & & & & & & & & & &\   &&&transport\\
\hline
 & &G&G&T&G&C&A& & &$\ $&-&heavy metal&siderophore\\
 & & & & & & & & & &\   & &ion porter& transport\\
\hline
 & &G&G&T&G&C& & & &\   &cell wall &-&-\\
 & & & & & & & & & &\   &(sensu Fungi)&-&-\\
\hline
{\bf A}&{\bf G}&{\bf G}&{\bf G}&{\bf T}&{\bf G}&{\bf C}&{\bf A}&{\bf C}&{\bf
C}&\multicolumn{4}{c|}{}\\
\hline
C&G&G&C&G&C&C& & & &\   &-&-&tricarboxylic\\
 & & & & & & & & & &\   &&&acid cycle\\
\hline
C&G&G&C&G&C&C&G& & &\   &-&-&tricarboxylic\\
 & & & & & & & & & &\   &&&acid cycle\\
\hline
 &G&G&C&G&C&C&G&A& &\   &-&-&tricarboxylic\\
 & & & & & & & & & &\   &&&acid cycle\\
\hline
 & &G&C&G&C&C&G&A&G&\   &-&-&tricarboxylic\\
 & & & & & & & & & &\   &&&acid cycle\\
\hline
{\bf C}&{\bf G}&{\bf G}&{\bf C}&{\bf G}&{\bf C}&{\bf C}&{\bf G}&{\bf A}&{\bf
G}&\multicolumn{4}{c|}{}\\
\hline
\end{tabular}
\end{center}

\end{table}

\newpage\noindent
\subsection*{Table 6}
Other significant motifs associated to transcription
factors not studied in Ref. \cite{lee:02}. \par \mbox{}

\begin{table}[h]
\begin{center}
\begin{tabular}{|c@{}c@{}c@{}c@{}c@{}c@{}c@{}c@{}c@{}c@{}c@{}|c|c|c|}
\hline
\multicolumn{11}{|c|}{motif} &C&F&P\\
\hline
A&C&C&C&C& & & & & &\   &alpha, alpha-trehalose-&-&-\\
 & & & & & & & & & &\   &phosphate synthase &&\\
 & & & & & & & & & &\   &complex (UDP-forming)&&\\
\hline
 &C&C&C&C&T& & & & &\   &alpha, alpha-trehalose-&-&-\\
 & & & & & & & & & &\   &phosphate synthase &&\\
 & & & & & & & & & &\   &complex (UDP-forming)&&\\
\hline
{\bf A}&{\bf C}&{\bf C}&{\bf C}&{\bf C}&{\bf T}&&&&&\multicolumn{4}{c|}{}\\
\hline
C&C&G&G&T&G&G&C& & &\   &26S proteasome&threonine&-\\
 & & & & & & & & & &\   &&endopeptidase&\\
\hline
 &A&G&G&T&G&G&C&A& &\   &26S proteasome&peptidase&-\\
\hline
 &C&G&G&T&G&G&C&A& &\   &26S proteasome&proteasome&-\\
 & & & & & & & & & &\   &&endopeptidase&\\
\hline
 &G&G&G&T&G&G&C&A& &\   &26S proteasome&proteasome&-\\
 & & & & & & & & & &\   &&endopeptidase&\\
\hline
{\bf C}&{\bf C}&{\bf G}&{\bf G}&{\bf T}&{\bf G}&{\bf G}&{\bf C}&{\bf
A}&&\multicolumn{4}{c|}{}\\
\hline
A&A&A&C&A&A&A& & & &\   &spindle&-&-\\
\hline
T&A&A&A&C&G&A& & & &\   &-&-&steroid metabolism\\
\hline
\end{tabular}
\end{center}

\end{table}

\newpage\noindent
\subsection*{Table 7}
Genome prevalence of the consensus sequences constructed by
aligning the motifs found significant by the algorithm, and functional
characterization of the corresponding genes. The second column
contains the total number of genes in the yeast genome in which the
consensus sequence is found at least once in the upstream region. The
third column reports one of the annotation terms to which the
consensus sequence is associated, and the number of genes annotated to
the term among the ones containing the consensus sequence. \par \mbox{}
\begin{table}[h]
\begin{center}
\begin{tabular}{|l|l|l|l|}
\hline
Consensus sequence&Total genes&Annotated genes\\
\hline
GATGAGATAGCT&17&nucleolus (8)\\
AAAAATTTTC&233&nucleolus (37)\\
MCSCMTACAY&78&cytosolic ribosome (20)\\
ARACGCGTMG&15&DNA replication and chromosome cycle (2)\\
AAMYGCGAWCG&6&nucleosome (0)\\
GATGAGTCAY&5&amino-acid metabolism (3)\\
TGAAACA&560&sexual reproduction (26)\\
AGGGTGCA&40&siderophore transport (5)\\
CGGCGCCGAG&5&tricarboxylic acid cycle (3)\\
ACCCCT&434&alpha, alpha-trehalose phosphate\\
&&synthase complex (UDP-forming) (2)\\
CCGGTGGC&17&26S proteasome (6)\\
\hline
\end{tabular}
\end{center}

\end{table}

\subsection*{Table 8}
Candidate targets of regulation by the UPC2p and ECM22p
  transcription factors. \par \mbox{}

\begin{table}[h]
\begin{center}
\begin{tabular}{|ll|}
\hline
ATF2&(YGR177C)\\
ECM22&(YLR228C)\\
ERG3&(YLR056W)\\
ERG4&(YGL012W)\\
ERG5&(YMR015C)\\
ERG7&(YHR072W)\\
ERG26&(YGL001C)\\
\hline
\end{tabular}
\end{center}

\end{table}

\newpage\noindent
\subsection*{Table 9}
Candidate targets of regulation by the Hcm1p
  transcription factor. \par \mbox{}

\begin{table}[h]
\begin{center}
\begin{tabular}{|ll|}
\hline
MPS1&(YDL028C)\\
CIN8&(YEL061C)\\
PDS1&(YDR113C)\\
SPC98&(YNL126W)\\
VIK1&(YPL253C)\\
SPC25&(YER018C)\\
ESP1&(YGR098C)\\
STU2&(YLR045C)\\
SLI15&(YBR156C)\\
\hline
\end{tabular}
\end{center}

\end{table}
\begin{figure}
  \caption{Location of the motifs belonging to the consensus GATGAGATGAGCT 
    with respect to the translation and transcription start sites for 6 genes 
    for which the latter is known. The binding sites are denoted by rectangles
    above or below the line depending on whether the consensus sequence is 
    read on the Crick or Watson strand respectively. The vertical bar is the 
    transcription start site, as given in Ref. [20].}
  \centering{\includegraphics[width=13.cm]{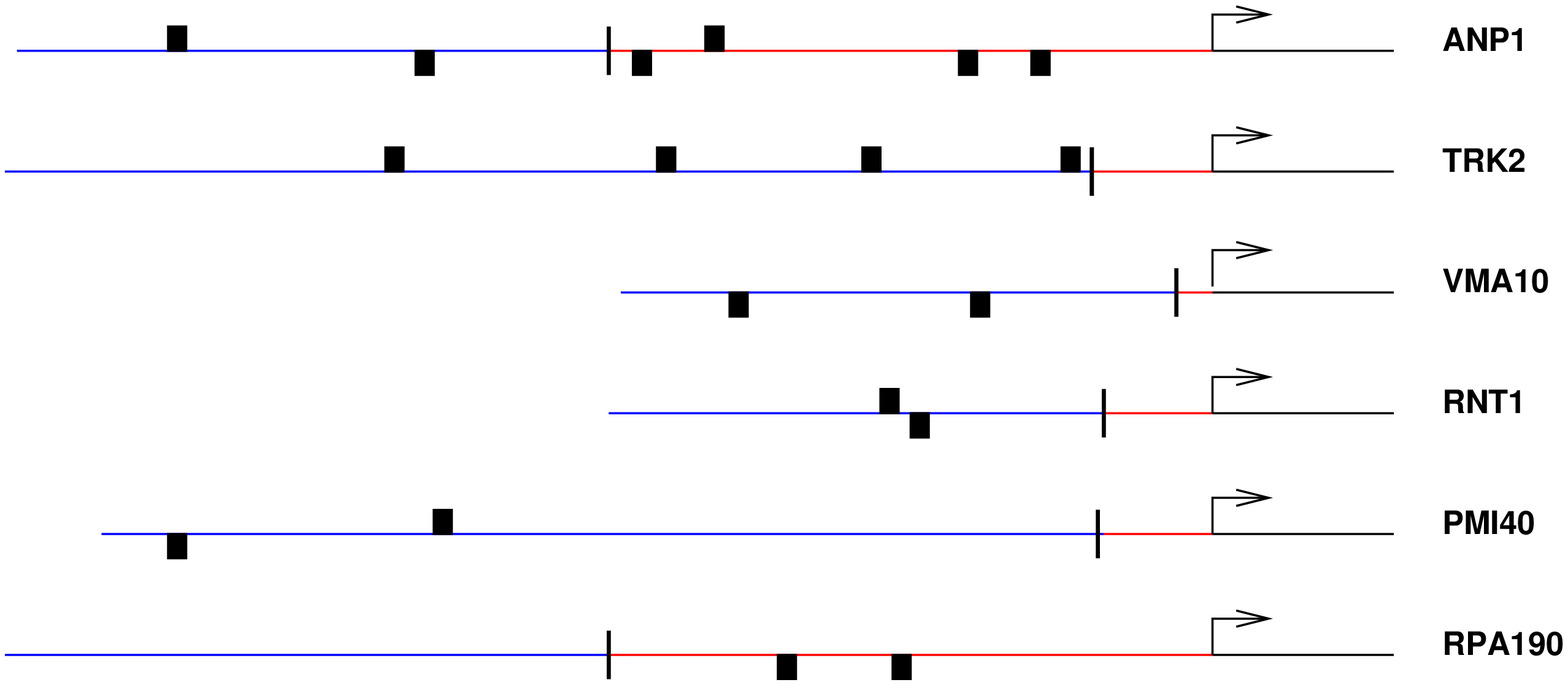}}
\end{figure}

\end{bmcformat}
\end{document}